\documentclass[final,twocolumn]{elsarticle}

\usepackage{hyperref}
\usepackage{amsmath}

\graphicspath{{figures/}}

\begin{document}

\begin{frontmatter}

\title{Monte Carlo calculation and verification of the geometrical factors for the NPDGamma experiment}

\author[utkaddress,ornladdress]{K. B. Grammer\corref{correspondingauthor}}
\cortext[correspondingauthor]{Corresponding author. Oak Ridge National Laboratory, PO BOX 2008 MS6466, Oak Ridge, TN 37831-6466}
\ead{grammerkb@ornl.gov; kgrammer@vols.utk.edu}

\author[asuaddress,anladdress]{D. Blyth}
\author[ornladdress]{J. D. Bowman}
\author[utkaddress]{N. Fomin}
\author[utkaddress,ornladdress]{G. L. Greene}
\author[utkaddress,mitaddress]{M. Musgrave}
\author[ornladdress,ukyaddress]{E. Tang}
\author[indaddress,lanladdress]{Z. Tang}

\address[utkaddress]{University of Tennessee, Knoxville, TN, USA}
\address[ornladdress]{Oak Ridge National Laboratory, Oak Ridge, TN, USA}
\address[asuaddress]{Arizona State University, Tempe, AZ, 85287, USA}
\address[anladdress]{Argonne National Laboratory, Argonne, IL, 60439, USA}
\address[mitaddress]{Massachusetts Institute of Technology, Cambridge, MA, 02139, USA}
\address[ukyaddress]{University of Kentucky, Lexington, Kentucky 40506, USA}
\address[indaddress]{Indiana University, Bloomington, IN 47405, USA}
\address[lanladdress]{Los Alamos National Laboratory, Los Alamos, NM 87545, USA}

\begin{abstract}
The NPDGamma experiment measures the parity-violating asymmetry in $\gamma$-ray emission in the capture of polarized neutrons on liquid parahydrogen. The sensitivity to the asymmetry for each detector in the array is used as a parameter in the extraction of the physics asymmetry from the measured data. The detector array is approximately cylindrically symmetric around the target and a step-wise sinusoidal function has been used for the sensitivity in the previous iteration of the NPDGamma experiment, but deviations from cylindrical symmetry necessitate the use of a Monte Carlo model to determine corrections to the geometrical factors. For the calculations, source code modifications to MCNPX were done in order to calculate the sensitivity of each cesium iodide detector to the physics asymmetry. We describe the MCNPX model and results from calculations and how the results are validated through measurement of the parity violating asymmetry of $\gamma$-rays from neutron capture on chlorine.\tnotetext[mytitlenote]{This manuscript has been authored by UT-Battelle, LLC under Contract No. DE-AC05-00OR22725 with the U.S. Department of Energy. The United States Government retains and the publisher, by accepting the article for publication, acknowledges that the United States Government retains a non-exclusive, paid-up, irrevocable, worldwide license to publish or reproduce the published form of this manuscript, or allow others to do so, for United States Government purposes. The Department of Energy will provide public access to these results of federally sponsored research in accordance with the DOE Public Access Plan (http://energy.gov/downloads/doe-public-access-plan).}\tnotetext[titlenote2]{\textsuperscript{\textcopyright} 2018. This manuscript version is made available under the CC-BY-NC-ND 4.0 license \href{http://creativecommons.org/licenses/by-nc-nd/4.0/}{http://creativecommons.org/licenses/by-nc-nd/4.0/}}
\end{abstract}

\begin{keyword}
Monte Carlo\sep MCNPX\sep NPDGamma experiment
\end{keyword}

\end{frontmatter}


\section{Introduction}
The goal of the NPDGamma experiment is to investigate the weak nucleon-nucleon interaction by measuring the parity violating asymmetry in the angular distribution of 2.2~MeV $\gamma$-rays emitted in the capture reaction
\begin{equation}
\vec{\textrm{n}}+\textrm{p}\to\textrm{d}+\gamma.
\end{equation}
The magnitude of the parity violating asymmetry is on the order of $10^{-8}$. The differential cross section for the capture reaction is proportional to the physics asymmetry, $A_\gamma$, and the angle between the neutron spin, $\hat{\sigma}_n$, and the $\gamma$-ray momentum, $\hat{k}_\gamma$,
\begin{equation}
\label{eq:npdg_diffxs}
\frac{d\sigma}{d\Omega}  \propto \frac{1}{4 \pi}(1+A_\gamma \hat{\sigma}_n \cdot \hat{k}_\gamma).
\end{equation}

\begin{figure}
\begin{center}
\includegraphics{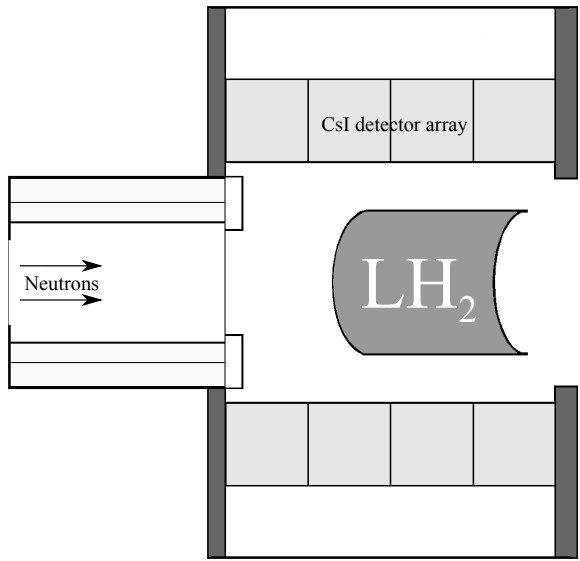}
\caption{Neutrons enter the RFSR from the left after passing through a collimator. The liquid hydrogen target is surrounded 48 CsI detectors. The beam propagates in the $+\hat{z}$ direction, the vertical direction is $+\hat{y}$, and beam left is the $+\hat{x}$ direction.}
\label{fig:model}
\end{center}
\end{figure}

\begin{figure}
\begin{center}
\includegraphics{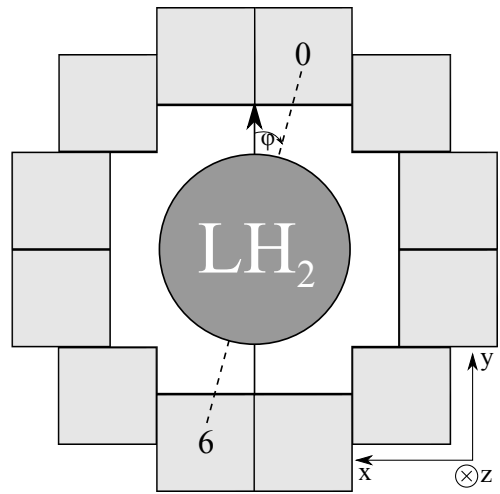}
\caption{Detector numbering scheme with the neutron propagation direction $(\hat{z}$) into the page. $\phi$ is measured relative to the vertical axis. Detector pairs are offset by $\phi=\frac{\pi}{2}$.}
\label{fig:array}
\end{center}
\end{figure}

The neutron beam is 10~cm x 12~cm at the exit of the Fundamental Neutron Physics Beamline (FNPB)~\cite{Fomin2015} at the Spallation Neutron Source at Oak Ridge National Laboratory. The beam passes through a supermirror polarizer~\cite{Balascuta2012} before entering the resonant frequency spin rotator (RFSR)~\cite{Seo2008}. The RFSR alternates the neutron spin over a sequence of 8 beam pulses in a $\uparrow\downarrow\downarrow\uparrow\downarrow\uparrow\uparrow\downarrow$ pattern in order to cancel first and second order fluctuations in the beam intensity. Finally, neutrons enter the NPDGamma detector array, which consists of 48 CsI(Tl) detectors~\cite{Gericke2005a} arranged as shown in figure \ref{fig:model}. Each detector consists of two 15.2~cm x 15.2~cm x 7.6~cm CsI(Tl) crystals. The detectors are arranged symmetrically around the neutron beam center in 4 rings of 12 surrounding the liquid hydrogen cryostat. The liquid hydrogen target is 30~cm long and 13~cm in radius~\cite{Santra2010}.
The $\hat{z}$ direction is the neutron beam direction, $\hat{x}$ direction is beam left, and the $\hat{y}$ direction is beam up. The neutron spin direction is $\pm\hat{y}$ depending on the state of the RFSR and the azimuthal angle, $\phi$, is the angle in the x-y plane (see figure \ref{fig:array}), and the polar angle, $\theta$, is the angle between the target and the each detector ring.

The raw asymmetry for a given spin sequence is determined using the super ratio method, in which we define
\begin{equation}
\alpha_i = \frac{N_{\uparrow}^{l}}{N_{\downarrow}^{l}} \frac{N_{\downarrow}^{m}}{N_{\uparrow}^{m}},
\end{equation}
for each pair, $i$, of opposed detectors, $l=0\ldots5$ and $m=l+6$. The detectors in each pair are separated by $180^\circ$ in $\phi$ (see figure \ref{fig:array}). The values of $N_{\uparrow}$ and $N_{\downarrow}$ are the spin up and spin down signals, respectively, in detectors $l$ and $m$. The raw asymmetry for each detector pair is then given by
\begin{equation}
A_{\textrm{raw}}^{i} = \frac{\sqrt{\alpha_i}-1}{\sqrt{\alpha_i}+1}.
\end{equation}
The physics asymmetry~\cite{Fry2017} for a given detector is given by
\begin{equation}
A_{\textrm{PC}} G_{\textrm{PC}}^{i} + A_{\textrm{PV}} G_{\textrm{PV}}^{i} = \frac{A_{\textrm{raw}}^{i} - A_{\textrm{app}}^{i}}{P_n \Delta_{\textrm{dep}} \Delta_{\textrm{sf}}},
\label{eq:npdg_physasym}
\end{equation}
where $A_{\textrm{PC}}$ and $A_{\textrm{PV}}$ are the parity-conserving and parity-violating physics asymmetries, $A_{\textrm{app}}^{i}$ is a term encompassing contributions that dilute the asymmetry, $P_n$ is the measured neutron beam polarization~\cite{Musgrave2018}, $\Delta_{\textrm{dep}}$ is the neutron depolarization correction, and $\Delta_{\textrm{sf}}$ is the spin flip efficiency. Using the super ratio method is superior to using the difference between detector signals because it perfectly accounts for the differences in the gains of the detectors in each pair and is less sensitive to differential detector pedestal values.

The pairwise geometrical factors $G_{\textrm{PC}}^{i}$ and $G_{\textrm{PV}}^{i}$ in equation \ref{eq:npdg_physasym} are related to the ideal single-detector geometrical factors, $G_{\textrm{LR}}^{j}$ and $G_{\textrm{UD}}^{j}$ discussed in the next section, which are proportional to the sensitivity of a detector pair to the physics asymmetry. The ideal geometrical factors are simply determined by the $(\theta,\phi)$ position of the detector relative to the source in the case of a point source and infinitely small detectors. The ideal geometrical factors depend on the geometry of the apparatus and beam and a further correction using the neutron momentum to generate $G_{\textrm{PC}}^{i}$ and $G_{\textrm{PV}}^{i}$ from the ideal geometrical factors, with the neutron momentum being used to properly account for neutrons that change direction before capture. The geometrical factors are determined using Monte Carlo and are used as fitting parameters in order to extract the physics asymmetry from the measured detector raw asymmetries.

Qualitatively, the shape of the sensitivity of the detector array to the physics asymmetry, $A_\gamma$, is expected to be sinusoidal function of $\phi$ within each detector ring with an additional dependence on the polar angle, $\theta$, for each detector ring relative to the neutron capture location. A step-wise sinusoidal function that did not take into account the polar angle for each ring was used for the geometrical factors in the analysis of the previous run of the NPDGamma experiment at the Los Alamos Neutron Science Center spallation source (LANSCE)~\cite{Gericke2011}. The calculation method described here is used to determine geometrical factors for all targets that have been used for the NPDGamma experiment, and this work describes a significant improvement to the determination of the geometrical factors over the LANSCE method. These targets include liquid hydrogen, $^{35}$Cl, aluminum target, and apparatus aluminum samples. Apparatus aluminum in the target vessel and elsewhere cause a dilution factor in the measurement of the hydrogen asymmetry. The measurement of the $^{35}$Cl asymmetry is used to validate the Monte Carlo calculation of the geometrical factors.

\section{MCNPX model for the ideal geometrical factors}
There are three geometrical factors that can be expressed in terms of the initial $\gamma$-ray direction, the neutron momentum, and the neutron spin and each can be expressed analytically for a simplified model with a point source and point detectors. The up-down geometrical factor is a projection of the $\gamma$-ray momentum along the y-axis in the assumption that the neutron spin is always parallel to the vertical axis,
\begin{equation}
G_{\textrm{UD}} = \langle\hat{k}_{\gamma}\cdot\hat{y}\rangle  = \langle\sin(\theta)\cos(\phi)\rangle.
\label{eq:gud_ideal}
\end{equation}
The left-right geometrical factor is a projection of the $\gamma$-ray momementum along the horizontal axis in the assumption that the neutron momentum is always parallel to the beam axis and neglecting scattering inside the target volume,
\begin{equation}
G_{\textrm{LR}} = \langle\hat{k}_{\gamma}\cdot\hat{x}\rangle =\langle\sin(\theta)\sin(\phi)\rangle.
\label{eq:glr_ideal}
\end{equation}
The beam axis geometrical factor is not used in the NPDGamma analysis and is left as a qualitative consistency check of the geometrical factors calculation,
\begin{equation}
G_{\textrm{Z}} = \langle\hat{k}_{\gamma}\cdot\hat{z}\rangle = \langle\cos(\theta)\rangle.
\label{eq:gz_ideal}
\end{equation}

The geometrical factors are sensitive to the alignment, position, and size of the beam, detector array, and target. A horizontal or vertical shift will lead to a change in the origin and therefore a change in $\phi$ resulting in significant change in the geometrical factors for individual detectors. However, pairwise geometrical factors that combine opposing pairs of detectors are insensitive to small shifts. For a pair of detectors,
\begin{equation}
\begin{array}{rcl}
G_{\textrm{UD}}^{l} & \approx & -G_{\textrm{UD}}^{m}, \\
G_{\textrm{LR}}^{l} & \approx & -G_{\textrm{LR}}^{m}. \\
\end{array}
\label{eq:pairsigns}
\end{equation}
When analyzing pairs of detectors, the pairwise geometrical factors are given as the difference of the individual detector geometrical factors,
\begin{equation}
\begin{array}{rcl}
G_{\textrm{UD}}^{i} & = & G_{\textrm{UD}}^{l}-G_{\textrm{UD}}^{m} \\
G_{\textrm{LR}}^{i} & = & G_{\textrm{LR}}^{l}-G_{\textrm{LR}}^{m}, \\
\end{array}
\label{eq:gpair}
\end{equation}
which are insensitive to small up-down and left-right shifts. For instance, take two point detectors, $l$ at coordinates $(15\textrm{~cm},15\textrm{~cm})$ and $m$ at $(-15\textrm{~cm},-15\textrm{~cm})$, with a point source at the origin. The geometrical factors for this configuration are $\pm \frac{\sqrt{2}}{2}$. Under a shift of 1~cm along the $\hat{x}$ (or $\hat{y}$) axis, the individual geometrical factors shift by $\approx$4\%, but the pairwise geometrical factors shift by less than 0.2\%.

The top and bottom detectors (see figure \ref{fig:array}) are most sensitive to an up-down asymmetry and therefore are expected to have the largest $G_{\textrm{UD}}$ and smallest $G_{\textrm{LR}}$, and the side detectors are expected to have the largest $G_{\textrm{LR}}$ and smallest $G_{\textrm{UD}}$. 

There are four corner detectors that are slightly closer to the hydrogen target but are also somewhat shielded by neighboring detectors. The detector array is symmetric under rotations by $\frac{\pi}{4}$ and $\frac{\pi}{2}$, which transforms $G_{\textrm{UD}} \to G_{\textrm{LR}}$ and $G_{\textrm{UD}} \to -G_{\textrm{UD}}$, respectively. The beam axis geometrical factor, $G_{\textrm{Z}}$, should be approximately constant within each ring, but geometrical factors for the corner detectors will be slightly shifted from the other 8 detectors.

The detectors are extended in space rather than point-like, and the center of detection may be perturbed from the geometric center because the detector package is comprised of 2 CsI(Tl) crystals gain matched to 10\%~\cite{Gericke2005a} with the seam between the crystals always parallel to the beam direction such that top detectors have ``beam left'' and ``beam right'' crystals rather than ``upstream'' and ``downstream'' crystals, for example. The detector array was aligned to the y and z-axes with an accuracy of 3~mrad. The nominal center of the detector array is 25~cm from the inward detector faces and the efficiency matching implies a shift of 0.75~cm in the center of detection which corresponds to an angular shift of 35~mrad. Finite geometry and shadowing effects also prevent an individual detector's $\phi$ and $\theta$ from being determined geometrically. In order to correct for these effects,  we built a Monte Carlo model using MCNPX 2.7~\cite{Pelowitz2011}. The MCNPX calculation uses a 10~cm x 12~cm neutron source beginning at the exit of the FnPB neutron guide several meters upstream of the detector array as well as an as-built model of the detector array and liquid hydrogen target apparatus. All material cross sections are taken from ENDF/B-VII.1~\cite{Chadwick2011}.

The energy deposition is tallied in each detector from $\gamma$-rays originating from neutron capture in the target along with information on the initial direction cosines for each $\gamma$-ray. The source code was modified to save the direction cosines for the initial momentum of a $\gamma$-ray in the model. As the $\gamma$-ray propagates through the model, these initial direction cosines remain unchanged and are also inherited by all new particles created throughout each source particle history, for instance electrons from Compton scattering. A $\gamma$-ray may produce Compton electrons in multiple detectors, which produces another deviation from the ideal sinusoidal function. As energy is deposited in the detectors by Compton scattered electrons, the initial $\gamma$-ray direction cosine tags are used to bin and weight the energy deposition tallies. The range of a 100~keV electron in CsI is only 60~$\mu$m~\cite{Berger2005}, and the propagation of electrons below 100~keV is terminated to save calculation time without compromising results.

The built in cell-averaged energy deposition tally (\texttt{F6}) in MCNPX was used to tally the energy deposition from neutron capture $\gamma$-rays in units of MeV g$^{-1}$ per source neutron. A \texttt{tallyx} subroutine was used to correctly weight and bin the energy deposition by the initial direction cosine tags. $E_{kl}^{j}$ represents the amount of energy deposited at each scattering event as determined by MCNPX as an electron scatters and loses energy. The energy deposition is summed over scattering events, $l$, and source neutron tracks, $k$. $N_n$ represents the total number of source neutrons generated for the calculation, and is the usual normalization factor used in MCNPX. A total of four tallies were calculated for each detector; an unmodified \texttt{F6} tally and three \texttt{F6} tallies weighted by the initial $\gamma$-ray direction cosine tags. The total energy deposition in each detector is determined using an unmodified \texttt{F6} tally and is given by,
\begin{equation}
T_{\textrm{ave}}^{j} = \frac{1}{N_n}\sum_{k=1}^{N_n}\sum_{l} E_{kl}^{j}.
\label{eq:tally_total}
\end{equation}

\begin{figure}[!h]
\centering
\includegraphics{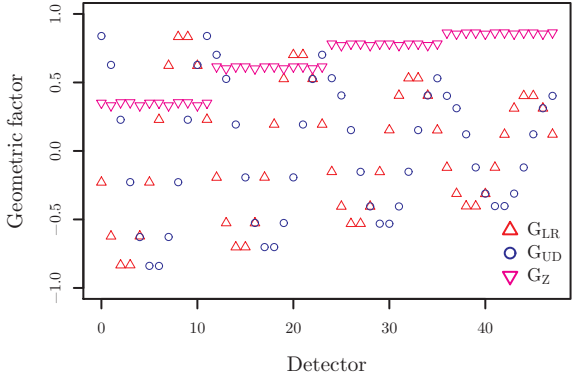}
\caption{$G_{\textrm{UD}}$, $G_{\textrm{LR}}$, and $G_{\textrm{Z}}$ ideal geometrical factors for the $^{35}$Cl target inside the downstream end of the RFSR.}
\label{fig:clsfgf_udlrz}
\end{figure}

The three directionally weighted energy deposition tallies are calculated identically to a normal , \texttt{F6} energy deposition tally but the energy deposition tally at each scattering event is weighted by the initial $\gamma$-ray direction tags using the \texttt{tallyx} subroutine, 
\begin{equation}
T_{x}^{j} = \frac{1}{N_n}\sum_{k=1}^{N_n}\sum_{l} E_{kl}^{j} (\hat{k}_{\gamma}\cdot\hat{x})_k,
\label{eq:tally_xw}
\end{equation}
\begin{equation}
T_{y}^{j} = \frac{1}{N_n}\sum_{k=1}^{N_n}\sum_{l} E_{kl}^{j} (\hat{k}_{\gamma}\cdot\hat{y})_k,
\label{eq:tally_yw}
\end{equation}
\begin{equation}
T_{z}^{j} = \frac{1}{N_n}\sum_{k=1}^{N_n}\sum_{l} E_{kl}^{j} (\hat{k}_{\gamma}\cdot\hat{z})_k.
\label{eq:tally_zw}
\end{equation}
The geometrical factors are calculated as the ratio of the direction weighted energy deposition to the total energy deposition for each detector,
\begin{equation}
G_{\textrm{LR}}^{j}=\langle\hat{k}_{\gamma}\cdot\hat{x}\rangle=\frac{T_{x}^{j}}{T_{\textrm{ave}}^{j}},
\label{eq:glr_tally}
\end{equation}
\begin{equation}
G_{\textrm{UD}}^{j}=\langle\hat{k}_{\gamma}\cdot\hat{y}\rangle=\frac{T_{y}^{j}}{T_{\textrm{ave}}^{j}},
\label{eq:gud_tally}
\end{equation}
\begin{equation}
G_{\textrm{Z}}^{j}=\langle\hat{k}_{\gamma}\cdot\hat{z}\rangle=\frac{T_{z}^{j}}{T_{\textrm{ave}}^{j}}.
\label{eq:gz_tally}
\end{equation}

Consider a source neutron that captures in the liquid hydrogen and produces a $\gamma$-ray that scatters in detectors $0$ and $1$, producing Compton electrons in each, and then escapes. In this situation, there will be contributions to the four $T^{0}$ and $T^{1}$ tallies due to the energy lost by those Compton electrons and zero contribution to any other detector. As another example, consider a source neutron that captures in the liquid hydrogen and produces a $\gamma$-ray that never enters a detector volume, in which case there will be no contribution to any detector tallies from this neutron source particle track.

Figure \ref{fig:clsfgf_udlrz} shows $G_{\textrm{LR}}$, $G_{\textrm{UD}}$, and $G_{\textrm{Z}}$ for the $^{35}$Cl target positioned inside the downstream end of the spin flipper. Qualitatively, the ring closest to the target has the highest sensitivity to the asymmetry, with each ring having a lower sensitivity as the polar angle decreases. The beam axis component is approximately constant for each ring of 12 detectors with only small deviations for each corner detector. There is also a $\frac{\pi}{4}$ phase between $G_{\textrm{LR}}$ and $G_{\textrm{UD}}$. 

\subsection{Neutron scattering correction}
Equations \ref{eq:gud_ideal} and \ref{eq:glr_ideal} include only the apparatus, detector, and neutron beam geometry contributions to the geometrical factors. The sensitivity to the parity conserving and parity violating asymmetries depends on the pseudoscalar quantity, $\hat{k}_{\gamma}\cdot\hat{\sigma}_n$, and scalar quantity, $\hat{k}_{\gamma}\cdot(\hat{\sigma}_n\times\hat{k}_n)$. The parity conserving geometrical factor is proportional to the parity conserving asymmetry, for instance Mott-Schwinger scattering~\cite{Gericke2008} and parity conserving apparatus asymmetry effects~\cite{Csoto1997}, and is perpendicular to the neutron spin,
\begin{equation}
G_{\textrm{PC}} = \langle\hat{k}_{\gamma}\cdot(\hat{\sigma}_n\times\hat{k}_n)\rangle,
\label{eq:gpc_ideal}
\end{equation}
while the parity violating asymmetry depends only on the neutron spin,
\begin{equation}
G_{\textrm{PV}} = \langle\hat{k}_{\gamma}\cdot\hat{\sigma}_n\rangle  = \langle\sin(\theta)\cos(\phi)\rangle.
\label{eq:gpv_ideal}
\end{equation}

In the case of a material with a large absorption cross section compared to scattering, neutrons are generally captured before scattering such that the initial momentum is unchanged and $G_{\textrm{PC}}$ would be the same as $G_{\textrm{LR}}$. The scattering cross section for parahydrogen is comparable to the absorption cross section for 5~meV neutrons, and neutron scattering must be taken into account when determining $G_{\textrm{PC}}$ such that the appropriate MCNPX tally is given by,
\begin{equation}
T_{\textrm{PC}}^{j} = \frac{1}{N_n}\sum_{k=1}^{N_n}\sum_{l} E_{kl}^{j} (\hat{k}_{\gamma}\cdot(\hat{\sigma}_n\times\hat{k}_n))_k,
\label{eq:tally_xw_pc}
\end{equation}
rather than equation \ref{eq:tally_xw}. When extracting the geometrical factor from the MCNPX tallies,
\begin{equation}
G_{\textrm{PC}}^{j}=\langle\hat{k}_{\gamma}\cdot(\hat{\sigma}_n\times\hat{k}_n)\rangle=\frac{T_{PC}^{j}}{T_{\textrm{ave}}^{j}}.
\label{eq:gpc_tally}
\end{equation}
The contribution due to the unknown amount of orthohydrogen contamination in the liquid hydrogen volume is assessed below. The neutron beam polarization and neutron depolarization correction on capture are taken into account separately as shown in equation \ref{eq:npdg_physasym}. The left-right and parity-conserving geometrical factors differ significantly because neutron scattering before capture is a significant contribution in the NPDGamma experiment. The neutron direction does not contribute to $G_{\textrm{PV}}$ in equation \ref{eq:gpv_ideal} and $G_{\textrm{PV}}$ is therefore taken to be the same as $G_{\textrm{UD}}$ and determined by equation \ref{eq:gud_tally}.

\section{Detector center of response}
\begin{figure}
\begin{center}
\includegraphics{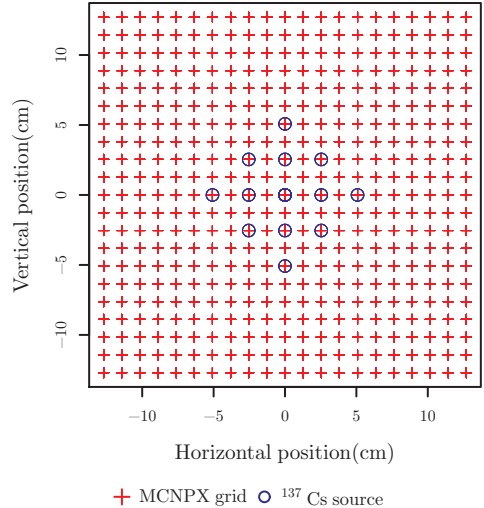}
\caption{Grid scan data points using the 4 mCi $^{137}$Cs source are shown in blue. The MCNPX model grid points are shown in red.}
\label{fig:grid}
\end{center}
\end{figure}

An 0.8~cm long, 0.3~cm radius cylindrical 4~mCi $^{137}$Cs source was translated using an x-y scanner along a 13 point grid in a plane centered within each ring of the detector array. Data were taken at each position for 60 seconds before moving the source to the next position in the grid. A much larger grid scan was simulated using MCNPX using a cylindrical $\gamma$-ray source with energy 662~keV in order to determine the ideal detector response. The grid scan patterns are shown in figure \ref{fig:grid}. The MCNPX grid represents the ideal response of the detectors to the cesium source and would differ from the source measurements by a scaling factor if not for imperfections in the crystals that give rise to spatial variations in the detector gain that we cast as an effective rotation of the detector array. These scans are used to determine these effective rotation angles, $\delta_\phi$, for each detector and then used to adjust the calculated geometrical factors as an admixture of $G_\textrm{PV}$ and $G_\textrm{PC}$. The detector package dimensions were chosen to be sufficiently thick to the 2.2~MeV $\gamma$-rays from capture on hydrogen and this also applies to the 662~keV $\gamma$-rays from $^{137}$Cs.

\begin{figure}
\centering
\includegraphics{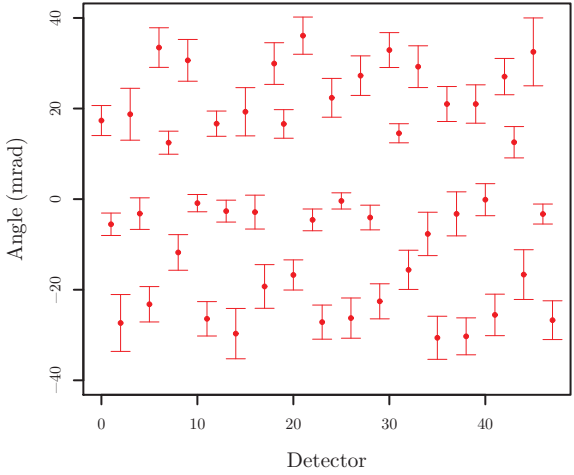}
\caption{Extracted $\delta_{\phi}$ correction angles with uncertainties determined from a fit of the measured detector response for a $^{137}$Cs source to the MCNPX calculated detector response using equation \ref{eq:gridfit}.}
\label{fig:delta}
\end{figure}

The MCNPX simulation results should be proportional to the average detector signal in volts for each detector along with a correcting rotation angle, $\delta_{\phi}$, that is expected to be on the order of 35~mrad. The ideal response function is a function of the position of the $^{137}$Cs source. This function is fit to an 8th order polynomial in x and y using GNU Scientific Library~\cite{Galassi},
\begin{equation}
P(x,y)=\sum_{p=0}^{8}\sum_{q=0}^{8-i} k_{p,q} x^p y^q.
\label{eq:gridfit}
\end{equation}
The laboratory coordinate system differs from the model coordinate system with a rotation by $\delta_{\phi}$,
\begin{equation}
\begin{pmatrix}
x' \\
y' 
\end{pmatrix} = 
\begin{pmatrix}
\cos(\delta_{\phi}) & -\sin(\delta_{\phi}) \\
\sin(\delta_{\phi}) & \cos(\delta_{\phi}) 
\end{pmatrix}
\begin{pmatrix}
x \\
y 
\end{pmatrix}.
\end{equation}
 The measured data, $S(x', y')$, is then fit to the model function along with a scaling factor, $b$, and a rotation angle, $\delta_{\phi}$,
\begin{equation}
S(x', y') = b P(x', y').
\end{equation}
The $\delta_{\phi}$ angles are within $\pm$40~mrad of the ideal system, as was expected. The uncertainties shown in figure~\ref{fig:delta} were taken from the fitting routine.

The adjusted geometrical factors are given by the simple transformation $\phi \rightarrow \phi + \delta_{\phi}$, which produces rather simple expressions for the adjusted geometrical factors given the complex geometry of the beam, target, and apparatus. The adjusted geometrical factors can then be expressed in terms of $\delta_{\phi}$ and the ideal $G_{\textrm{PC}}$ and $G_{\textrm{PV}}$,
\begin{equation}
\begin{array}{rcl}
G_{\textrm{PC}}' & = & \langle\hat{k}_{\gamma}\cdot\hat{x}\rangle' \\
 & = & \langle\sin(\theta)\sin(\phi+\delta_{\phi})\rangle \\
 & = & \langle\hat{k}_{\gamma}\cdot\hat{x}\rangle\cos(\delta_{\phi})+\langle\hat{k}_{\gamma}\cdot\hat{y}\rangle\sin(\delta_{\phi}) \\
G_{\textrm{PC}}' & = & G_{\textrm{PC}}\cos(\delta_{\phi})+G_{\textrm{PV}}\sin(\delta_{\phi}). \\
\end{array}
\label{eq:kxprime}
\end{equation}
Similarly, the adjusted up-down geometrical factor is given by,
\begin{equation}
G_{\textrm{PV}}'=G_{\textrm{PV}}\cos(\delta_{\phi})-G_{\textrm{PC}}\sin(\delta_{\phi}).
\label{eq:kyprime}
\end{equation}
The $\delta_{\phi}$ adjustment contributes an uncertainty on the order of 0.5\% for the most statistically significant detectors (ie. top/bottom detectors for $G_\textrm{PV}$) and on the order of 2\% for the least statistically significant detectors (ie. side detectors for $G_\textrm{PV}$). It is the corrected $G_{\textrm{PC}}'$ and $G_{\textrm{PV}}'$ that are used in the asymmetry analysis.

\section{Validation of MCNPX results with $^{35}$Cl parity-violating asymmetry}
$^{35}$Cl has a large and well known parity-violating asymmetry of $(2.91 \pm 0.67) \times 10^{-5}$~\cite{Fomin2012sq} and can be measured at FNPB in approximately 1 day of neutron beam time, compared to the hydrogen asymmetry which is approximately 3 orders of magnitude smaller. Previous measurements have have shown that the parity-conserving asymmetry for $^{35}$Cl is consistent with zero at the level of $7 \times 10^{-6}$~\cite{Mitchell2004}. A dimensionally thin $^{35}$Cl target that is opaque to neutrons was used in order to validate the calculation of the geometrical factors as well as to test the sensitivity of the apparatus to a parity-violating signal. The chlorine target consists of a Teflon case filled with carbon tetrachloride. The target was placed inside the downstream end of the RSFR (see figure \ref{fig:model}) in order to perform periodic checks on the apparatus.
\begin{figure}[!h]
\begin{center}
\includegraphics{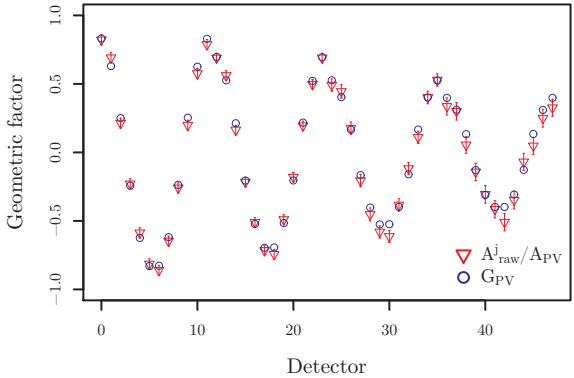}
\caption{$^{35}$Cl asymmetry fitted to ring-specific geometrical factors (ie. amplitude for each ring varying with $\cos(\theta)$) and scaled by the resulting scaling parameter, $A_{\textrm{PV}}$, in equation \ref{eq:npdg_physasym2}. The reduced $\chi^2$ is 0.88.}
\label{fig:clasym_w_geo}
\end{center}
\end{figure}
\begin{figure}[!h]
\begin{center}
\includegraphics{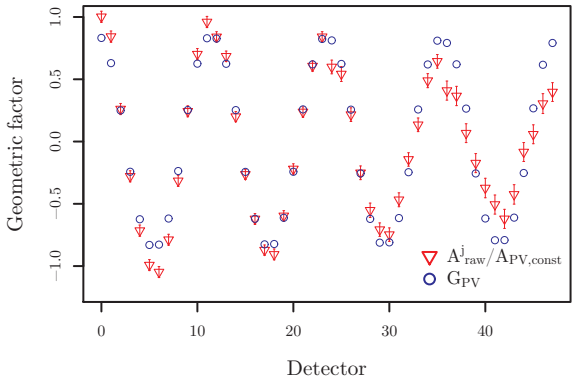}
\caption{$^{35}$Cl asymmetry fitted to constant amplitude geometrical factors for each ring (analogous to the LANSCE procedure~\cite{Gericke2011}) and scaled by the resulting scaling parameter, $A_{\textrm{PV,const}}$, in equation \ref{eq:npdg_physasym2}. The reduced $\chi^2$ is 2.57.}
\label{fig:clasym_w_geo_constant}
\end{center}
\end{figure}

The raw per-pulse detector $\gamma$-ray response signals were cleaned using simple cuts that ensure that the neutron beam was stable throughout a spin sequence, and the resulting asymmetry was recorded for each spin sequence that satisfied these cuts. The spin sequence asymmetries were then recorded in a histogram and fit to a Gaussian in order to extract a raw asymmetry and an uncertainty for each detector, which was then fit to the geometrical factors using
\begin{equation}
A_{\textrm{raw}}^{j} = A_{\textrm{PC}} {G_{\textrm{PC}}^{j}}' + A_{\textrm{PV}} {G_{\textrm{PV}}^{j}}'.
\label{eq:npdg_physasym2}
\end{equation}

The $\chi^2$ per degree of freedom from this fit is 0.88 when using the adjusted geometrical factors described above. The PV geometrical factors are shown in figure \ref{fig:clasym_w_geo} along with the raw asymmetries with the asymmetry fit parameter, $A_{\textrm{PV}}$, divided out so that the asymmetries appear on a scale of -1 to 1. The fit shows good agreement between the predicted shape of the asymmetry (from $G_{\textrm{PV}}$) and the measured detector asymmetries, with the amplitude decreasing with each ring as the angle $\theta$ increases from rings 1 to 4.

The same fitting procedure was performed using the ring 1 geometrical factors applied to every ring, which is analogous to the LANSCE procedure~\cite{Gericke2011} in which a constant amplitude function was used for each ring. The $\chi^2$ per degree of freedom is 2.57 using this method, and the fit is visually quite poor and can be seen in figure \ref{fig:clasym_w_geo_constant} with the parameter $A_{\textrm{PV,const}}$ divided out, which demonstrates the validity of the Monte Carlo calculation that takes into account both the $\phi$ and $\theta$ angles. The uncertainty budget for the geometrical factors for the chlorine target is shown in table \ref{tab:uncertainty}. Simulations accounting for $\pm$0.5cm longitudinal misalignment of the chlorine target provide an additional uncertainty of 0.5\% in the geometrical factors.

\begin{table}[!h]
\begin{center}
    \begin{tabular}{| l | c |}
    \hline
    Source & Uncertainty \\ \hline
    Calculation & 0.2\% \\
    $\delta_{\phi}$ adjustment & 0.5\% \\ 
    Alignment & 0.5\% \\ \hline
    Total & 0.7\% \\ \hline
    \end{tabular}
\caption[Table caption text]{Geometrical factors uncertainty budget for $^{35}$Cl.}
\label{tab:uncertainty}
\end{center}
\end{table}

\section{Liquid hydrogen target}
The hydrogen target vessel contains a mixture of orthohydrogen and parahydrogen. The target was operated at 15.6~K, at which the equilibrium orthohydrogen concentration is $0.015\%$. There is a recirculation loop through the ortho-para converter (OPC)~\cite{Barron-Palos2011} that drives conversion towards the parahydrogen ground state. The time constant for conversion was determined from neutron transmission measurements to be approximately 1 day. It was not possible to determine the absolute orthohydrogen concentration during the experiment via neutron transmission and there was no in situ apparatus for measuring the orthohydrogen concentration directly. However, we performed measurement of the parahydrogen scattering cross section~\cite{Grammer2015} using the NPDGamma apparatus that yielded upper and lower bounds on the orthohydrogen concentration of 0.15\% and 0.015\%, respectively.

\begin{figure}[!h]
\centering
\includegraphics{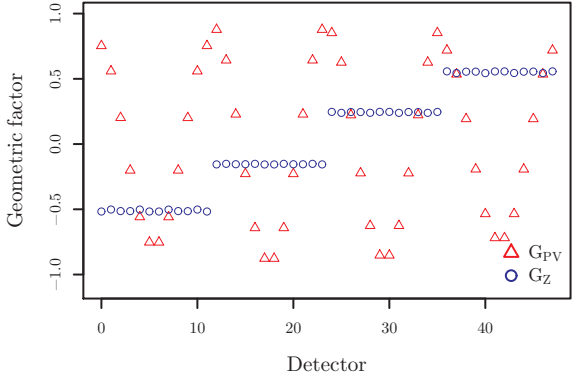}
\caption{$G_{\textrm{PV}}$ and $G_{\textrm{Z}}$ ideal geometrical factors for the liquid hydrogen target.}
\label{fig:hydroarray_gfuz}
\end{figure}
\begin{figure}[!h]
\centering
\includegraphics{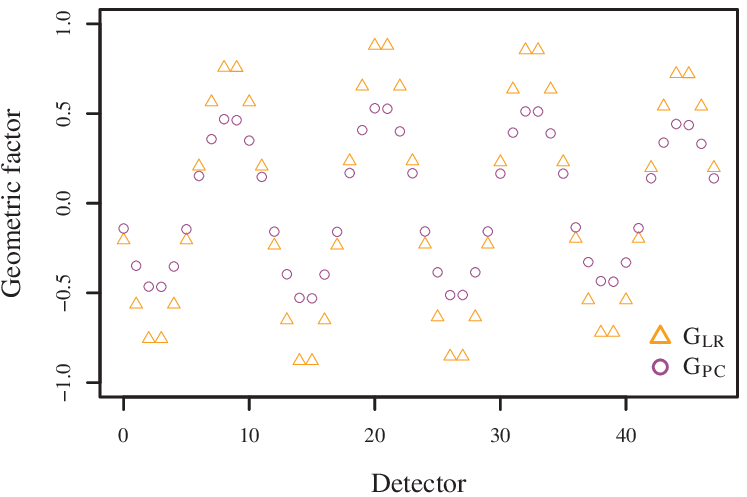}
\caption{$G_{\textrm{LR}}$ and $G_{\textrm{PC}}$ ideal geometrical factors for the liquid hydrogen target.}
\label{fig:hydroarray_gflr}
\end{figure}
\begin{figure}[!h]
\centering
\includegraphics{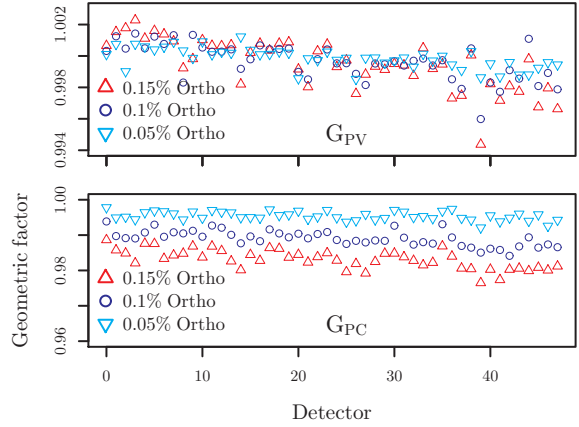}
\caption{Ratio of the PV (top) and PC (bottom) geometrical factors for varying orthohydrogen concentrations relative to the lower bound orthohydrogen concentration (0.015\%).}
\label{fig:ortho_dep_gf}
\end{figure}
The liquid hydrogen geometrical factors are shown in figures \ref{fig:hydroarray_gfuz} and \ref{fig:hydroarray_gflr}. The neutron scattering correction to $G_{\textrm{LR}}$ is responsible for the significant difference between $G_{\textrm{LR}}$ and $G_{\textrm{PC}}$.

The orthohydrogen scattering cross section is a factor of $10^2$ higher in magnitude than the absorption and parahydrogen scattering cross sections~\cite{Chadwick2011}\cite{mughabghab2006atlas}. The mean free path of neutrons is sensitive to the orthohydrogen concentration, and the orthohydrogen concentration dependence for the PV and PC geometrical factors is shown in figure \ref{fig:ortho_dep_gf}, which contributes a $1.0\%$ uncertainty to the hydrogen geometrical factors. There is also an energy dependence to the geometrical factors due to the energy dependence of the mean free path of a neutron in liquid hydrogen, which contributes an uncertainty of $1.5\%$. The uncertainties in the geometrical factors for the liquid hydrogen target are shown in table \ref{tab:uncertainty_hydro}. The effect of  misalignment along the beam axis of $\pm$1~cm was calculated in the case of the hydrogen target. The position of the target along the beam axis relative to the detector array changes $\cos{(\theta)}$ and therefore the absolute magnitude of the sensitivity. The $G_\textrm{Z}$ geometrical factor has a very strong sensitivity to beam axis shifts and changes by 5\% to 15\% depending on the ring. The $G_\textrm{LR}$ and $G_\textrm{UD}$ geometrical factors are less sensitive to this shift, with rings 2 and 3 changing by 0.5\% and rings 1 and 4 changing by 1.5\%.
\begin{table}
\begin{center}
    \begin{tabular}{| l | c |}
    \hline
    Source & Uncertainty \\ \hline
    Calculation & 0.2\% \\
    $\delta_{\phi}$ adjustment & 0.5\% \\ 
    Energy dependence & 1.5\% \\
    Ortho-Para & 1.0\% \\
    Alignment & 0.5\% \\ \hline
    Total & 1.9\% \\ \hline
    \end{tabular}
\caption[Table caption text]{Geometrical factors uncertainty budget for hydrogen.}
\label{tab:uncertainty_hydro}
\end{center}
\end{table}

\section{Conclusion}
We have shown that the Monte Carlo method using MCNPX to calculate the geometrical factors can be used to determine geometrical factors for our target configurations. This was done to a precision of approximately 1.9\% for the distributed liquid hydrogen target and 0.7\% for the $^{35}$Cl target used in the NPDGamma experiment. The uncertainty in the calculated geometrical factors makes a negligible contribution to the overall uncertainty in the NPDGamma physics asymmetry.

\section{Acknowledgements}
This work was supported in part by DOE DE-FG02-03ER41258, NSF PHY-0457219, and NSF PHY-0758018. Z. Tang acknowledges support from the Indiana University Center for Spacetime Symmetries.

\bibliography{papers_writeups-geofactors.bib}

\end{document}